\documentclass{sig-alternate-2013}
\usepackage[utf8x]{inputenc}
\usepackage{amsfonts}
\usepackage{amsmath}
\usepackage{clrscode3e}
\usepackage{graphicx}
\usepackage{hyperref}
\usepackage{relsize}
\usepackage{tabularx}
\usepackage{varwidth}
\usepackage{wrapfig}

\makeatletter
\def\@copyrightspace{\relax}
\makeatother

\permission{Copyright is held by the International World Wide Web Conference Committee (IW3C2). IW3C2 reserves the right to provide a hyperlink to the author's site if the Material is used in electronic media.}
\conferenceinfo{WWW 2015,}{May 18--22, 2015, Florence, Italy.}
\copyrightetc{ACM \the\acmcopyr}
\crdata{978-1-4503-3469-3/15/05. \\
http://dx.doi.org/10.1145/2736277.2741111}

\usepackage[
  pass,
]{geometry}

\begin{document}

\title{Parallel Streaming Signature EM-tree: \\
A Clustering Algorithm for Web Scale Applications}
\numberofauthors{2}
\author{
\alignauthor
Christopher M. De Vries \\
       \affaddr{Game Analytics ApS}\\
       \affaddr{Berlin, Germany}\\
       \email{cdevries@gameanalytics.com}
\alignauthor
Lance De Vine\\
       \affaddr{Queensland University of Technology}\\
       \affaddr{Brisbane, Australia}\\
       \email{l.devine@hdr.qut.edu.au}
\and
\alignauthor
Shlomo Geva\\
       \affaddr{Queensland University of Technology}\\
       \affaddr{Brisbane, Australia}\\
       \email{s.geva@qut.edu.au}
\alignauthor
Richi Nayak\\
       \affaddr{Queensland University of Technology}\\
       \affaddr{Brisbane, Australia}\\
       \email{r.nayak@qut.edu.au}
}

\date{10 November 2015}

\maketitle
\begin{abstract}

The proliferation of the web presents an unsolved problem of automatically analyzing billions of pages of natural language. We introduce a scalable algorithm that clusters hundreds of millions of web pages into hundreds of thousands of clusters. It does this on a single mid-range machine using efficient algorithms and compressed document representations. It is applied to two web-scale crawls covering tens of terabytes. ClueWeb09 and ClueWeb12 contain 500 and 733 million web pages and were clustered into 500,000 to 700,000 clusters.  To the best of our knowledge, such fine grained clustering has not been previously demonstrated. Previous  approaches clustered a sample that limits the maximum number of discoverable clusters.  The proposed EM-tree algorithm uses the entire collection in clustering and produces several orders of magnitude more clusters than the existing algorithms.  Fine grained clustering is necessary for meaningful clustering in massive collections where the number of distinct topics grows linearly with collection size. These fine-grained clusters show an improved cluster quality when assessed  with two novel evaluations using ad hoc search relevance judgments and spam classifications for external validation. These evaluations solve the problem of assessing the quality of clusters where categorical labeling is unavailable and unfeasible.
\end{abstract}




\newpage
\section{Introduction}

Clustering is a fundamental process for applications such as Content Analysis, Information Integration, Information Retrieval, Web Mining and Knowledge Discovery. The proliferation of the internet has driven the need for unsupervised document clustering for analyzing natural language without having to label all possible topics such as in supervised learning approaches. Other web data exists with the potential of generating millions of clusters from billions of examples. For example, sensor data, images, video, audio, customer data, smart grid data, and, data produced by everyday physical objects such as in the web of things. Clustering algorithms are essential for these emerging web applications, and it is challenging due to heavy computational requirements. It may be possible to achieve web-scale clustering using a high performance distributed architecture, but the cost is prohibitive for many applications.  Additionally, high-performance computing platforms are  limited to organizations with large budgets and highly skilled employees. Additionally, achieving fine-grained clustering poses difficulties even with high-performance architectures. We found no examples of clustering near a billion natural language documents.

Information Retrieval utilizes document clustering for pre-clustering collections on to multiple machines for distributed search. Collection distribution uses clustering for organization into thematic groups. These clusters are ranked by collection selection to determine which clusters to search. Collection selection selects a few relevant thematically related clusters for each query and therefore improves search performance. Only the top k documents that are returned by the initial search are considered for further analysis when ranking search results.  By contrast, global pre-clustering of documents captures the thematic association of documents on the basis of shared content.  Documents that may not contain the search terms, but are thematically related to documents that are initially retrieved by keywords search, can be found in re-ranking and query expansion. Finding hidden thematic relationships based on the initial keyword search provides added motivation to pre-clustering document collections.

We are not aware of any solutions to the problem of clustering a billion web pages on standard hardware, other than by sampling to produce a relatively small number of clusters.  It is not possible to cluster a collection into numerous small clusters with only a small sample.  Clustering algorithms require redundancy in the data, and a small number of documents in the sample means reduced redundancy;  this does not allow for discovery of small clusters.  The size of the final clusters produced by sampling is relatively large because of distribution of the entire collection over the learned clusters. For instance, Kulkarni and Callan \cite{Kulkarni2010} have clustered a 0.1\% sample of ClueWeb09 into 1,000 clusters and then mapped all 500 million documents onto these clusters, yielding an average cluster size of 500,000 documents.  These large clusters work for collection distribution and selection, but we later demonstrate the advantage of finer grain clusters.  Furthermore, coarse clusters are not useful for topical clustering of documents, where topic cluster size is expected to be several of orders of magnitude smaller.  Fine grained clustering is not achievable through aggressive subsampling in web-scale collections.

This paper introduces the parallel streaming EM-tree algorithm for clustering web-scale collections on low-cost standard hardware.  The quality of clustering solutions and processing efficiency depend on the representation of objects.  Therefore, we present a useful data model using random projections \cite{Geva2011} for representation of the document collection as binary signatures.  We apply the parallel streaming EM-tree clustering algorithm that is specialized for binary signatures to segment the collection into fine grained clusters.  The proposed method is evaluated with the web scale collections, ClueWeb09\footnote{\url{http://ktree.sf.net/emtree/clueweb09}} and ClueWeb12\footnote{\url{http://ktree.sf.net/emtree/clueweb12}}, containing 500 and 733 million English language documents respectively.

Document clusters are often evaluated by comparison to a ground truth set of categories for documents. No topical labels are available for ClueWeb, and it is nigh impossible to generate the labels for a large scale web crawl for millions of categories. We present two novel methods for external cluster validation: (1) ad hoc relevance; and (2) spam classification. We used relevance judgements from the TREC Web Track in 2010, 2011, 2012 and 2013 \cite{Clarke2010,Clarke2011,Clarke2012,Clarke2013} and spam classifications created by Cormack et. al. \cite{Cormack2011} for the evaluation of document clustering.

Extensive analysis reveals that the clusterings with 500,000 to 700,000 clusters were found to improve the quality using this evaluation.  We emphasize that there are no earlier reports in the open literature of document clustering of this magnitude, and there are no standard benchmark resources or comparative evaluation results.

This paper makes two novel contributions: (1) we introduce the novel parallel streaming signature EM-tree algorithm that can cluster documents at scales not previously reported; and (2) we solve the problem of cluster validation in web-scale collections where creating a ground truth set of categories is a near impossible task for human assessors.

Section \ref{sec:related} discusses related research on web scale clustering.  The generation of a document representation is discussed in Section \ref{sec:representation}. Section \ref{sec:complexity} discusses the complexity of the proposed method and section \ref{sec:emtree} introduces the EM-tree algorithm. The evaluation of cluster validity is presented in Section \ref{sec:evaluation}. The discussion of potential applications and conclusions are contained in Sections \ref{sec:applications} and \ref{sec:conclusion}.

\section{Related Research}
\label{sec:related}

Few clustering algorithms can scale to web collections on modest hardware platforms. They rely on supercomputing resources or imposed constraints such as: (1) using a small sample for clustering and then mapping the entire collection to these clusters; (2) by reducing the number of target clusters.

MapReduce has been used to implement clustering algorithms  by Jin \cite{Jin2013} and Kumar \cite{Kumar2013}. The experiments created a small number of clusters using small data sets several gigabytes in size.  Esteves and Rong  \cite{Esteves2011} investigated using the open source Mahout library implemented using MapReduce for the clustering of Wikipedia. The k-means algorithm took two days to cluster the 3.5 million Wikipedia articles into 20 clusters using ten compute nodes. In comparison, we increase {\em both} the number of documents and clusters by two to three orders of magnitude. We analyze the complexity of k-means and EM-tree in Section \ref{sec:complexity}.

Broder et.  al. \cite{Broder2014} describe an approach to scaling up k-means by the use of inverted indexes. It reverses the process of assigning points to clusters and assign clusters to points. The authors also highlight the need for fine-grained clustering of high dimensional data such as web pages, users, and, advertisements. They indicate that k-means may take thousands of hours to converge when using parallel and distributed programming techniques such as MapReduce. While the authors show improvements to k-means, they only experiment with clustering millions of examples into thousands of clusters. We far exceed this scale.

Bahmani et. al. \cite{Bahmani2012} present a scalable parallel approach to the $D^2$ seeding approach from k-means++ \cite{Arthur2007} called k-means||. The experiments run k-means to complete convergence after initialization. It is often impractical to run the optimization to complete convergence. Typically, more than 95\% of the optimization happens in the first few iterations, and further iterations offer very little improvement.  We observed this with EM-tree in prior work \cite{DeVriesPhD}. The EM-tree can be seeded with k-means||. However, for fine-grained clustering of ClueWeb using signatures we found no advantage to more computationally expensive initializations. It may be due to the smoothing properties of random projections, but analysis of this is beyond the scope of this paper. However, when such an approach does help, our approach is complementary to k-means||. We can leverage the advantages of both approaches by the use of the scalable initialization procedure and the scalable optimization of EM-tree. As with other approaches, the experiments were run on data sets orders of magnitude smaller and much lower dimensionality.

Zhang et. al. \cite{Zhang1996} present the BIRCH algorithm that incrementally constructs a tree as data is streamed from disk. It is similar to the K-tree algorithm except that not all data points are kept in the tree \cite{DeVries2009a}. The BIRCH algorithm performs updates along the insertion path like K-tree. When using signatures it has been reported to cause scalability problems because the signature bits are continually unpacked and repacked \cite{DeVriesPhD}. Furthermore, the EM-tree has immutable tree state at each iteration leading to scalable parallel implementations. We were unable to find any reference to BIRCH clustering near 1 billion documents.

We have only found a few other examples of clustering near a billion examples.  These reports were on lower dimensional data sets of images, image patches, low dimensional generic point sets and weather patterns.  Liu et.  al.  \cite{Liu2007} used 2000 CPUs, MapReduce and commodity hardware to cluster 1.5 billion images into 1 million clusters for near duplicate detection.  Similarly to our approach, they used random projections and a tree structure.  Wang et.  al.  \cite{wang2013duplicate} also describe an approach to clustering images for duplicate detection using a 2000 core cluster for clustering 2 billion images. Bisgin and Dalfes  \cite{Bisgin2008} used 16,000 CPUs and a top500 supercomputer to cluster weather data into 1000 clusters.  Wu et. al.  \cite{wu2009clustering} report on the speedup obtained when using GPUs instead of multicore CPUs, to cluster billions of low dimensional point sets of upto 8 dimensions. These approaches rely on high-performance computing resources that put it out of reach for most application developers. Additionally, we found no examples of clustering web-scale document collections into fine-grained clusters.

The ClueWeb collections are some of the largest document collections used for research. These web crawls have been used at TREC to evaluate ad hoc retrieval and other Web search tasks. While it is clear that retrieval systems can scale to these collections, there has been little investigation of clustering such large document collections.

\section{Document Representation with Signatures}
\label{sec:representation}

The first task in clustering documents is the definition of a representation.  We use a binary signature representation called TopSig\footnote{\url{http://topsig.googlecode.com}}  \cite{Geva2011}. It offers a scalable approach to the construction of document signatures by applying random indexing  \cite{Sahlgren2005}, or random projections  \cite{Bingham2001} and numeric quantization. Signatures derived in this manner have been shown to be competitive with state of the art retrieval models at early precision, and also to clustering approaches.
The binary signature vectors faithfully preserve the mutual similarity relationships between documents in the original representation.  The Johnson Lindenstrauss lemma  \cite{Johnson1984} states that if points in a high dimensional space are projected into a randomly chosen subspace then the distances between the points are approximately preserved. The lower dimensionality required is asymptotically logarithmic with respect to the original high dimensional space.

This signature generation is similar to that of SimHash  \cite{Charikar2002}.  However, we use a different term weighting scheme, use signatures more than an order of magnitude larger, and, much sparser random codes.  SimHash has predominantly been applied to nearest duplicate detection where relatively short signatures are used to find the few nearest neighbors of a document. We refer the reader to  \cite{Geva2011} for the specific details of the signature generation process.

We removed stop words, and stemmed words using the Porter algorithm. We use 4096-bit signatures because that choice was previously shown to be sufficient to produce the same quality clustering as the original real-valued representation of documents \cite{Geva2011}. We derived the signatures used to represent the ClueWeb collections on a single 16 core machine. Our approach uses a fixed amount of memory while indexing and can process a collection without keeping the entire index in memory. Each document is indexed independently of all other documents leading to massive parallelization.

It is important to understand that the focus of this paper is efficient, scalable document clustering, not to compare different approaches of the generation of document signatures.  There have been numerous approaches reported that fall under the general category of similarity preserving hashing \cite{Wang2014}.

The use of signatures is advantageous for increased computational efficiency of document to document similarity. It has been shown to provide a one to two orders of magnitude increase in processing speed for document clustering over traditional sparse vector representations with sacrificing quality \cite{Geva2011}. In prior research \cite{Geva2011}, we used an algorithm similar to k-means designed especially for signatures. We specifically designed the proposed EM-tree algorithm for signatures. All documents and cluster representatives are binary signatures.  Therefore, traditional vector space clustering approaches can not be applied to these signatures without expanding them into a larger representation such as integer or floating point vectors.

\section{Clustering with EM-tree}
\label{sec:emtree}

The EM-tree algorithm can cluster vast collections into numerous fine-grained clusters. K-means is one of the earliest and most popular clustering algorithms due to its quick convergence and linear time complexity.  However, it is not suitable for web scale collections that have a vast diversity of content resulting in an enormous number of topics as elaborated in the previous section. EM-tree solves this previously unaddressed issue.

\subsection{M-Way Nearest Neighbor Search Tree}

A $m$-way nearest neighbor search tree is a recursive data structure that indexes a set of $n$ binary signatures in $d$ dimensions where, $X = \{x_1, \dots, x_n\}$, $X \subset \{+1,-1\}^d$ and $|X| = n$.  Each node contains a list of $record$s that are $(key, value)$ pairs. The $key$s are binary signatures that are cluster representatives of the associated subtree. Each iteration of the optimization updates the $key$s.  The $record$ values are the nodes associated with the $key$s that are one level deeper. The data structure also generalizes to all vector space representations in $\mathbb{R}^d$. However, in the case of the EM-tree, we have specialized the data structure for binary signatures. It is not a traditional vector space clustering algorithm.

Applying the k-means clustering algorithm recursively is known as Tree-Structured Vector Quantization (TSVQ) \cite{Gersho1993}, repeated k-means, or for clusters of size two, bisecting k-means. It generates a $m$-way tree in $\mathbb{R}^d$.

\subsection{The EM-tree Algorithm}

The EM-tree algorithm iteratively optimizes a randomly initialized $m$-way tree until convergence. In contrast, the repeated k-means algorithm initially creates $k$ clusters using k-means \cite{Gersho1993}. It recursively clusters each partition in a layer wise fashion until reaching a desired tree depth or node size.

The EM-tree builds a cluster tree in a different manner \cite{DeVriesPhD}. The collection is initially partitioned by selecting a random set of data points as cluster prototypes. Unlike repeated k-means, clustering is not applied at this point. Instead, the data points are recursively distributed to random partitions until a desirable tree depth is reached. At this point, the initial tree is complete. Now cluster means are updated in a bottom-up fashion. The entire process of tree insertion and tree update forms an iteration of the optimization and has been proven to converge \cite{DeVriesPhD}. The entire tree is recomputed with each insert update cycle. It is different from repeated k-means in which the optimization process is run to completion at each node before proceeding deeper into the tree. The EM-tree algorithm includes an additional pruning step in for removing empty branches of the tree. The EM-tree is not a standard vector space clustering algorithm in this context. It works directly with binary vectors where all documents and cluster prototypes are binary vectors.

The EM-tree algorithm can optimize any $m$-way tree such as a tree produced by a low-cost algorithm with poor cluster quality. The algorithm can be applied to any subtree in a $m$-way tree. In a setting where a changing data set is being clustered, branches of the tree affected by insertions and deletions can be restructured to the data independently of the rest of the tree.

By a process of insertion (Expectation), update (Maximization), and pruning, the $m$-way tree model adapts to the underlying data as the clusters converge as seen in Figure \ref{emtree_pseudo}.

The procedure \proc{seed} initializes the EM-tree algorithm where $m$ is the tree order, $X$ is the set of data points to cluster and $depth$ is the tree depth. It produces a height-balanced tree where all leaves are at the same depth.
The \proc{insert} procedure inserts a set of vectors into a $m$-way tree. Points are inserted by following the nearest neighbor search path, where at each node in the tree, the branch with the nearest key is followed.

The \proc{update} procedure updates the means in the tree according to the current assignment of data points in the leaves. Since we work with binary vectors, the bits in each signature assigned to a given leaf node are unpacked and accumulated into an integer vector. This vector is then used in two different ways. It is quantized to a binary vector to form a new cluster mean for a given leaf node, and it is also propagated up the tree so that new cluster centroids higher in the tree can be computed. The updating of cluster centroids is performed for all levels of the tree.

The \proc{prune} procedure removes any branches with no associated data points. It is completed bottom up where leaf nodes are removed first. The empty branches are removed once \proc{update} and \proc{insert} have completed. It allows the tree structure to adapt to the data.

The optimization error of EM-tree is robust with respect to different initializations when producing fine-grained clusters of ClueWeb. A tree with random initialization and subsequent optimization was found to minimize the objective function on par with more computationally expensive approaches. A 10\% sample is used to seed the tree because it is large enough to produce meaningful centroid representations for a large number of clusters. This initial seeding only consumes a small percentage of the compute time. Additionally, approximation guarantees can be achieved by initializing with the $D^2$ approach as in k-means++ \cite{Arthur2007} but analysis of this is beyond the scope of this paper.

In summary, the \proc{emtree} procedure initializes and iteratively optimizes an EM-tree of order $m$ that is $depth$ levels deep.

\begin{figure}
\begin{center}
\begin{varwidth}{\columnwidth}
\begin{codebox}
\Procname{$\proc{emtree}(m, depth, X)$}
\li $root = \proc{seed}(m, depth, X)$
\li $converged$ = false
\li \While not $converged$
\li \Do
    $root^\prime = root$
\li $\proc{insert}(root^\prime, X)$
\li $\proc{prune}(root^\prime)$
\li $\proc{update}(root^\prime)$
\li \If $root == root^\prime$
\li \Then
    $converged$ = true
\li \Else
\li $root = root^\prime$
    \End
    \End
\li \Return $root$
\end{codebox}
\end{varwidth}
\end{center}
\caption{EM-tree}
\label{emtree_pseudo}
\end{figure}

\subsection{Streaming EM-tree}

One of the key constraints in scaling clustering algorithms to large data sets is the availability of computer memory. One way to approach this problem is to adopt a streaming paradigm in which the data points are streamed sequentially, and only a small portion of them are ever kept in memory at one time. The approach used in the streaming EM-tree is only to keep internal nodes in memory. The data points are collected in accumulators associated with each leaf node. The data points inserted are, therefore, added into the accumulators but are then discarded.

\begin{figure}
\centering
\includegraphics[width=\columnwidth]{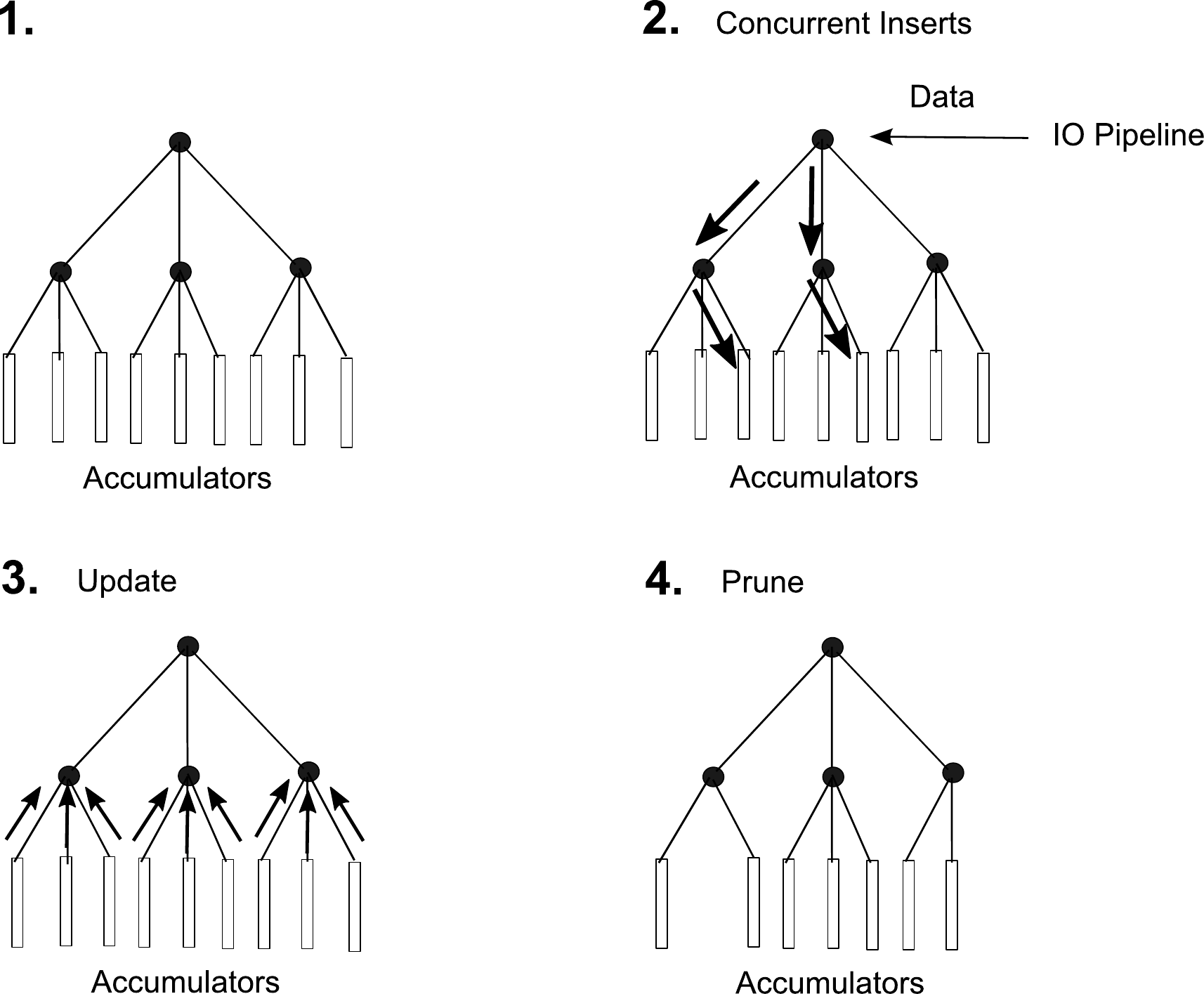}
\caption{Streaming EM-tree Iteration}
\label{fig:StreamingEM}
\end{figure}

At each iteration of the optimization, all signatures are read from disk and inserted into the tree as seen in Figure \ref{fig:StreamingEM}. Bits are unpacked from the signature and added into the accumulators. A count is kept of the number of points added into the accumulator. When all inserts have been performed, the values of the accumulators in the leaf nodes are propagated up the tree and new centroids are calculated in the \proc{update} step. The centroids, which are now vectors of integers, are quantized to produce new bit signatures. The \proc{prune} step is then performed.

Large datasets can be clustered on a single machine with limited memory. The algorithm is compute bound on typical hardware architectures the data is streamed from disk.

\newpage
\subsection{Parallelizing EM-tree}

The EM-tree algorithm is particularly amenable to large scale parallel and distributed implementations due to the nature of the optimization process. The key property that ensures scalability is that the entire tree is immutable at each iteration, and the data points can be concurrently inserted according to the nearest neighbor search path.

The EM-tree algorithm has been parallelized using threads via the use of loop parallelization and producer-consumer pipelines. The accumulators in the leaf nodes are locked when updated, although the chance of multiple threads attempting to update the same accumulator at the same time is small due to the typically very large number of leaf nodes. This property ensures that EM-tree scales almost linearly with regards to the number of threads. For thread implementation, we use Intel’s Threading Building Blocks (TBB).

A parallel threaded implementation can be executed on a single machine in which case the entire tree structure with accumulators is shared between threads. The bottleneck for the streaming EM-tree is the insertion step, primarily because of nearest neighbor calculations and bit unpacking of accumulators. The \proc{update} and \proc{prune} steps require much less computation.

\begin{figure}
\centering
\includegraphics[width=\columnwidth]{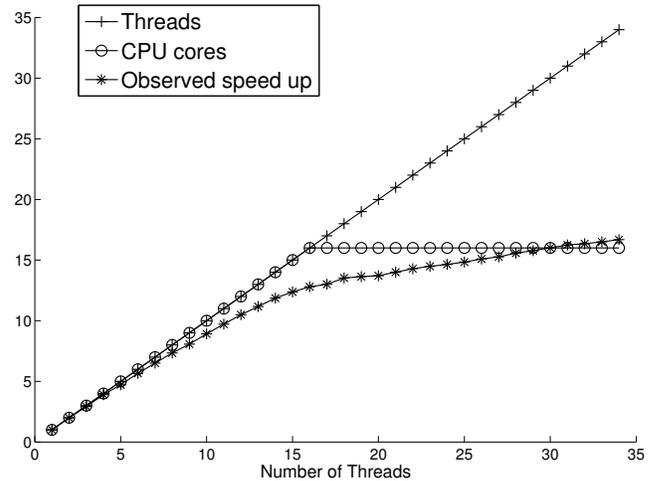}
\caption{Scaling EM-tree to 16 CPU cores}
\label{fig:scaling}
\end{figure}

Figure \ref{fig:scaling} contains the results of an experiment to test how EM-tree scales with the number of threads used. We tested EM-tree on a machine with 16 CPU cores. The speedup is measured relative to the execution in a single thread on the same hardware. The CPUs support two hyper-threads per core, and we have stepped through the number of threads in the program from 1 up to 34. The top curve in Figure \ref{fig:scaling} shows the speedup that could be obtained if speedup exactly followed the number of threads. It is not achievable in practice when using hyper-threads. The bottom curve shows the actual execution speedup as the number of threads is increased. The middle curve depicts the number of cores used -- when there are fewer threads than cores the threads are mapped to idle cores. After that point, all CPU cores are utilized, and threads are mapped to busy cores using their second hyper thread. In summary, on a machine with 16 cores we were able to achieve a 16 fold speedup through conventional multi-threading. In this experiment, we used a 10 million document sample from the ClueWeb09 collection. It is large enough to eliminate any caching effects and small enough that the single threaded experiment does not take weeks to run.

\subsection{Software}

EM-tree and Streaming EM-tree have been implemented by the authors as part of the open source LMW-tree C++ software package\footnote{\url{http://github.com/cmdevries/LMW-tree}}.  The software has been built on Linux, Mac OS and Windows. The clusters and tree structure have been made available online\footnote{\url{http://sf.net/projects/ktree/files/clueweb_clusters/}}. It is important to have these openly available since any alternative solution can be compared.

\section{Complexity Analysis}
\label{sec:complexity}

The time complexity of the k-means algorithm is linear with respect to all its inputs, $O(nki)$ \cite{Broder2014}.  Where $n$ is the number of non-zero entries in the document-term matrix, $k$ is the number of clusters, and, $i$ is the number of iterations.  The number of iterations, $i$, is limited to some small fixed amount of iterations as k-means converges quickly in practice. Fine grained clustering pushes $k$ much closer to $n$. As $k \rightarrow n$, the complexity of k-means approaches $O(n^2i)$. It makes k-means impractical for fine-grained clustering of web-scale collections containing billions of examples. Algorithms using height balanced trees such as EM-tree, K-tree \cite{DeVries2009a} and BIRCH \cite{Zhang1996} alleviate this problem by reducing the time complexity associated with the number of clusters to $O(\log(k))$. The complexity of EM-tree approaches $O(n \log(n) i)$ as $k \rightarrow n$ for fine-grained clustering.

Three nontrivial computer architecture considerations contribute to the efficiency of our approach. We process 64 dimensions of the binary signatures at a time in a single CPU operation. While it is a constant speedup, it is certainly not negligible. Our approach only uses integer operations that are faster and have higher throughput in instructions per cycle on modern processors than floating point operations used for k-means clustering.  Furthermore, due to the extremely concise representation of our model as binary signatures,  we observe more cache hits.  Cache misses are extremely costly on modern processors with main memory accesses taking more than 100 cycles.  When using traditional sparse representations, the cluster means in the root of the tree contain many millions of terms, quickly exceeding processor cache size. With signatures a cluster means always consumes 4096 bits.

By the use of efficient representations and algorithms we turn the problem of fine grained clustering of the entire web into a tractable problem.

\newpage
\section{Empirical Analysis}
\label{sec:evaluation}

We ran the experiments on a dual-socket Xeon E5-2665 based system. Each CPU package has $8 \times 2.4$ GHz CPU cores connected in a ring topology, for a total of 16 CPU cores. The program used 64GB of memory to store the tree in memory where the data set is 240GB to 350GB on disk. We streamed the data points from a 7200rpm 3TB SATA disk providing 150MB per second of sequential read performance, although this much bandwidth was not needed. A mid-range machine like this can be purchased for around 5,000 USD.

The ClueWeb 09 and 12 collections took approximately two and three days to index and resulted in 240GB and 350GB signatures indexes. It is a concise representation of the collections that measure in tens of terabytes when uncompressed. The EM-tree processes these document signatures to produce document clusters. Clustering the ClueWeb 09 and 12 signatures took approximately 15 and 20 hours. Both ran for five iterations and produced 700,000 and 600,000 clusters respectively.

Most of the literature on cluster evaluation focuses on evaluating document clusters by comparison to the ground truth set of categories for documents. It poses problems when evaluating large-scale collections containing millions to billions of documents. Human assessors are required to label the entire collection into many thousands of potential topics. Even if a small percentage of the collection is labeled, how does an assessor choose between many thousands of potential topics in a general purpose document collection such as the World Wide Web? The experience of labeling large test collections such as RCV1 \cite{Lewis2004} demonstrate that categorizing a general purpose document collection such as the web is a daunting task for humans. RCV1 contains 800,000 short newswire articles with hundreds of categories. Even though this collection is many orders of magnitude simpler than the web, assessors still struggled with categorization.

There are no topical human generated labels available for ClueWeb. Therefore, we propose novel uses of external sources of information as proxies for cluster validation: (1) ad hoc relevance; and (2) spam classification. The information retrieval evaluation community has already dealt with the problem of assessor load in ad hoc relevance evaluation via the use of pooling \cite{Voorhees2002}. Pooling reduces assessor load and topics evaluated are specific and well defined. The issue of assessor load is alleviated by using ad hoc relevance to assess document clustering, instead of category labels for every document in a collection.

We also present another alternative evaluation based upon spam classification produced in earlier research by Cormack et. al. \cite{Cormack2011}. The goal in this case is to place documents with the same spam score into the same cluster. It measures how consistent the clustering is with respect to the external measure of spam learned from 4 different spam classifications. We of course conjecture that there is some underlying topical vocabulary that spammy documents use. Clustering, grouping documents on the basis of shared vocabulary, should show a correlation with spam detection scores -- which are, of course, independently derived.

Before we present results of these two cluster validation experiments, we present some qualitative evaluation by inspection of partial documents content in the final trees in Tables \ref{table:meetup.com}, \ref{table:diseases} and \ref{table:infotech}.
\begin{center}
\begin{table*}[ht]
{\small
\hfill{}
\begin{tabularx}{\linewidth}{XXXXX}
\hline\noalign{\smallskip}
\textbf{Weight Loss} & \textbf{Investment} & \textbf{Language}
& \textbf{Politics} & \textbf{SciFi and Fantasy} \\
9f08ef2b0 & 9f06c7770 & 9f0476410 & 9f0a77b10 & 9f0b871e0 \\
\scriptsize{220 documents} & \scriptsize{1169 documents}
& \scriptsize{106 documents} & \scriptsize{1213 documents}
& \scriptsize{198 documents} \\
\noalign{\smallskip}
\hline
\noalign{\smallskip}
\noalign{\smallskip}
383 bits -- find a meetup group near you weight loss meetups pelham &
463 bits -- find a meetup group real estate buying investing meetups
alexandria &
386 bits -- find a meetup group linguistics meetups aurora &
637 bits -- find a meetup group democratic underground meetups portland &
667 bits -- find a meetup group star trek meetups long beach \\
 & & & & \\
417 bits -- find a meetup group weight loss meetups englewood &
587 bits -- find a meetup group real estate buying investing meetups decatur &
630 bits -- find a meetup group near you language lovers meetups naperville &
851 bits -- find a meetup group democratic underground meetups worcester &
706 bits -- find a meetup group star trek meetups allendale \\
 & & & & \\
474 bits -- find a meetup group fitness meetups staten island &
801 bits -- find a meetup group investing meetups lake jackson &
941 bits -- find a meetup group near you scandinavian languages meetups skokie &
1097 bits -- find a meetup group near you dennis kucinich meetups lakewood &
1009 bits -- find a meetup group near you comic books meetups reading \\
 & & & & \\
550 bits -- find a meetup group fitness meetups new rochelle &
998 bits -- find a meetup group near you real estate buying investing meetups
soulsbyville &
1252 bits -- find a meetup group near you hungarian language meetups
libertyville     &
1671 bits -- harford county democrats&
1157 bits -- find a meetup group buffy meetups london \\
 & & & & \\
580 bits -- find a meetup group weight loss meetups clinton &
1752 bits -- westfield group mergers and acquisitions alacrastorecom &
1844 bits -- algebra 1 terms and practice problems chapter 1 lesson 1 flash
cards quizlet &
1911 bits -- welsh statutory instruments 2005 &
1804 bits -- fleet registration station archive star trek online forums \\
\noalign{\smallskip}
\hline
\noalign{\smallskip}
\end{tabularx}
}
\hfill{}
\centering{\url{http://ktree.sourceforge.net/emtree/clueweb09/9f033ca50.html}}
\caption{ClueWeb09 -- Sample From Meetup.com Cluster}
\label{table:meetup.com}
\end{table*}
\end{center}

\begin{center}
\begin{table*}[ht]
{\small
\hfill{}
\begin{tabularx}{\linewidth}{XXXXX}
\hline\noalign{\smallskip}
\textbf{Hepatitis} & \textbf{HIV Vaccine} & \textbf{Treating HIV}
& \textbf{Disease Research} & \textbf{Bacteria} \\
a5b5f0820 & a5ae2f420 & a5b62af80 & a5b48c900 & a5b88ca20 \\
\scriptsize{774 documents} & \scriptsize{165 documents}
& \scriptsize{752 documents} & \scriptsize{288 documents}
& \scriptsize{126 documents} \\
\noalign{\smallskip}
\hline
\noalign{\smallskip}
\end{tabularx}
}
\hfill{}
\centering{\url{http://ktree.sourceforge.net/emtree/clueweb09/a5acc38a0.html}}
\caption{ClueWeb09 -- Sample From Diseases Clusters}
\label{table:diseases}
\end{table*}
\end{center}

\begin{center}
\begin{table*}[ht]
{\small
\hfill{}
\begin{tabularx}{\linewidth}{XXXXX}
\hline\noalign{\smallskip}
\textbf{Logistics Software} & \textbf{Video Surveillance}
& \textbf{Computer Security} & \textbf{Wireless}
& \textbf{Enterprise Software} \\
a7aded0e0 & a7b3b8f60 & a7b145970 & a7b5417c0 & a7b361450 \\
\scriptsize{213 documents} & \scriptsize{818 documents}
& \scriptsize{2007 documents} & \scriptsize{326 documents}
& \scriptsize{415 documents} \\
\noalign{\smallskip}
\hline
\noalign{\smallskip}
\end{tabularx}
}
\hfill{}
\centering{\url{http://ktree.sourceforge.net/emtree/clueweb09/a7aa623c0.html}}
\caption{ClueWeb09 -- Sample From Information Technology Cluster}
\label{table:infotech}
\end{table*}
\end{center}

\subsection{Ad Hoc Relevance Based Evaluation}
\label{sec:relevance}

The use of ad hoc relevance judgments for evaluation of document clusters is motivated by the cluster hypothesis \cite{Jardine1971}. It states that relevant documents tend to be more similar to each other than non-relevant documents for a given query. The cluster hypothesis connects ad hoc information retrieval and document clustering. Documents in the same cluster behave similarly with respect to the same information need \cite{Manning2008,Nayak2010,DeVries2011}. It is due to the clustering algorithm grouping documents with similar vocabularies together. There may also be some higher order correlation due to effects caused by the distributional hypothesis \cite{Harris1954} and limitation of the analysis to the size of the document.

\subsubsection {Oracle Collection Selection}

We have evaluated document clustering solutions by creating plots for the optimal ordering of clusters for ad hoc queries. This ordering is created by an {\em oracle collection selection} ranking that has full knowledge of relevant documents. Clusters are ordered by the number of relevant documents they contain. Cumulative recall is calculated by traversing the cluster ranking in descending order. Then the percentage of recall and documents visited is averaged across all queries. It represents the optimal ordering of the given clusters for a given query and represents an upper bound on recall ranking performance for the given set of clusters. Relevant documents group together in clusters when the cluster hypothesis holds. Additionally, this grouping is better than expected from randomly distributing documents into clusters of the same size. In summary, we assign to any clustering solution the performance of the oracle in collection selection. The better the clustering solution, the better the oracle performs in collection selection. Fewer clusters have to be searched to recall relevant documents.

There are no baselines available for clustering the ClueWeb collections into the order of a million clusters. Therefore, we compare our clustering solutions against a random partitioning baseline. A common problem when comparing different clustering solutions is that, by nature, different clustering solutions are produced by different methods. It can introduce structural bias to the evaluation. Unequal cluster sizes can lead to different performance measurements. In order to compare any given clustering solution against an equivalent random partitioning, we impose the same cluster structure on the random solution. We assign documents randomly to the same cluster structure obtained from clustering. In this manner, there is no structural bias to either solution, random or derived through clustering \cite{DeVries2012}. It eliminates advantages of ineffective clustering solutions. For example, if almost all documents are placed in a single cluster except for one document being placed in every other cluster, then the large cluster will almost always contain all the relevant documents to any query. An evaluation that tests to see if collection selection discovers the correct cluster to search concludes that the dysfunctional solution is indeed the best -- it almost always finds the right cluster. On the other hand, if the same ineffective solution is compared to a random baseline having the same dysfunctional structure then the clustering solution it is found to offer no improvement. The performance of different clustering solutions is normalized against the performance of a random baseline solution having an identical clustering structure.

\subsubsection{ClueWeb Results}

For the ClueWeb09 collection, we have evaluated three different clustering solutions. The first approach is by Kulkarni and Callan \cite{Kulkarni2010} using KL-divergence and k-means and the other two by the EM-tree algorithm. First we describe the approaches, and then compare their quality.

Kulkarni and Callan describe an experiment with ClueWeb09 where a 500,000 document sample of the ClueWeb09 collection was clustered using k-means with a KL-divergence similarity measure to produce 1000 clusters. It maps the 500 million documents onto these clusters. We have chosen this analysis for a direct comparison with EM-tree, even though it does not represent fine-grained clustering. However, it allows us to perform a side by side comparison with a published baseline. We processed the entire ClueWeb with an EM-tree containing two levels of tree nodes of order $m = 1000$. The first level contained 1000 clusters and the second 691,708 clusters due to pruning. The first level of the EM-tree and the method of Kulkarni and Callan using k-means both produced 1000 clusters. Therefore, these clustering solutions are directly comparable. We also compare solution quality of fine grained clusters, using the second level of the EM-tree. We demonstrate that using many more fine-grained clusters can produce higher quality clusters with respect to ad hoc relevance. Furthermore, the EM-tree can produce higher quality clusters with respect to spam, even at 1000 clusters. Direct comparison KL-divergence k-means approach is not possible for 691,708 clusters. It does not scale to produce this many clusters.

\begin{figure*}
\begin{minipage}[b]{0.33\linewidth}
\includegraphics[width=\columnwidth]{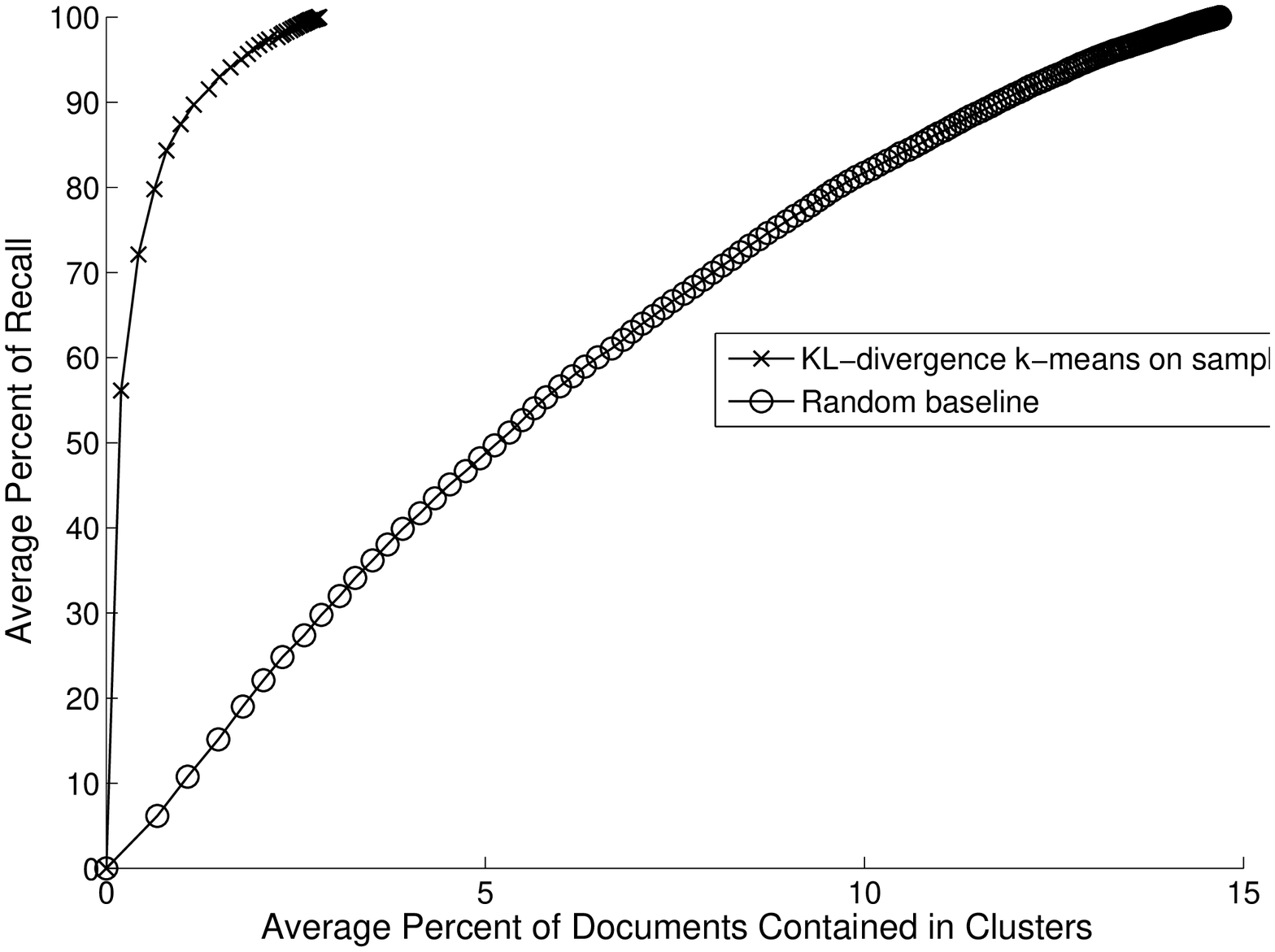}
\caption{ClueWeb09 -- 1000 KL-divergence k-means clusters}
\label{fig:kulkarni_callan_1000}
\end{minipage}
\begin{minipage}[b]{0.33\linewidth}
\includegraphics[width=\columnwidth]{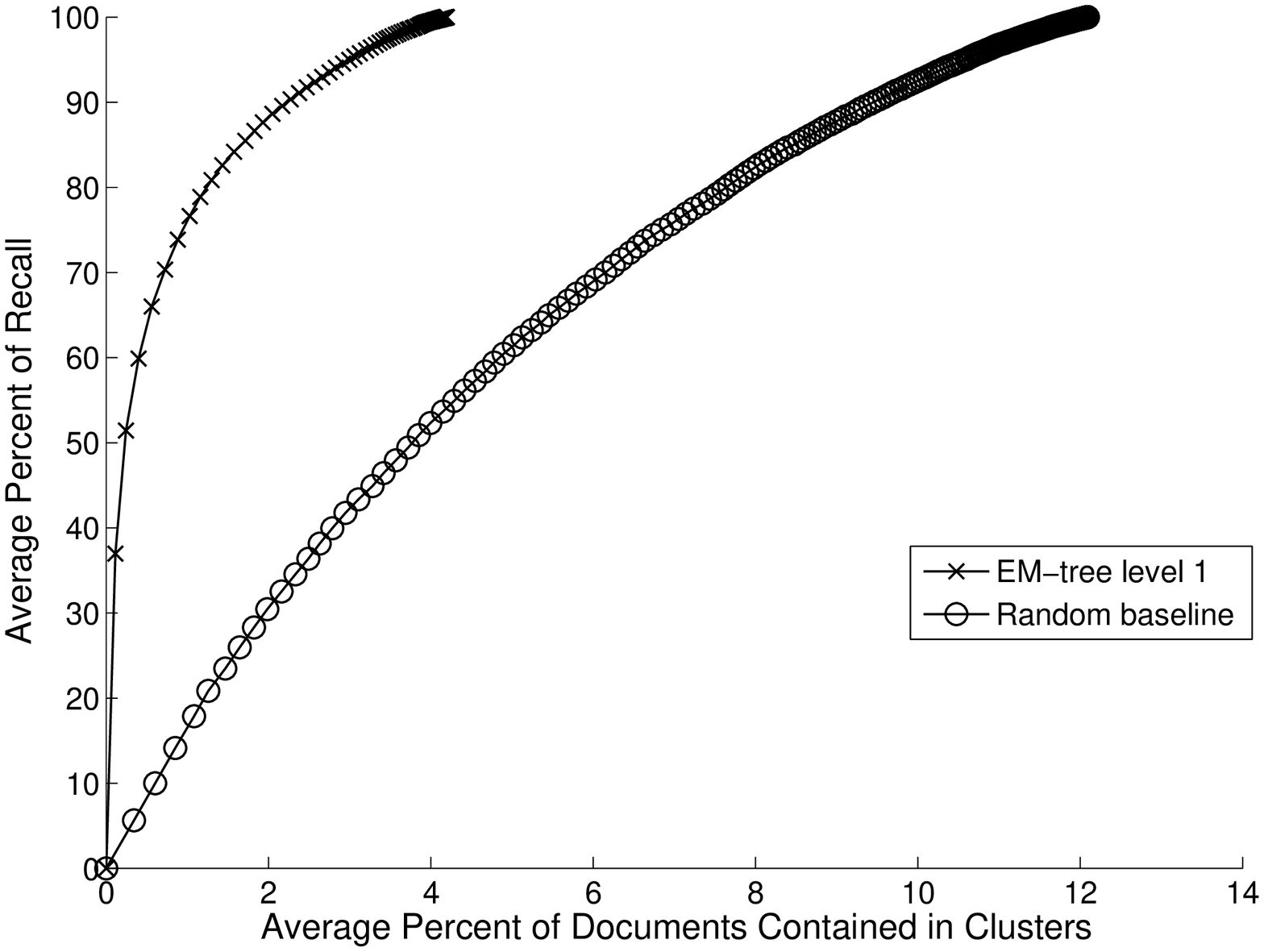}
\caption{ClueWeb09 -- 1000 EM-tree level 1 clusters}
\label{fig:emtree_1000}
\end{minipage}
\begin{minipage}[b]{0.33\linewidth}
\includegraphics[width=\columnwidth]{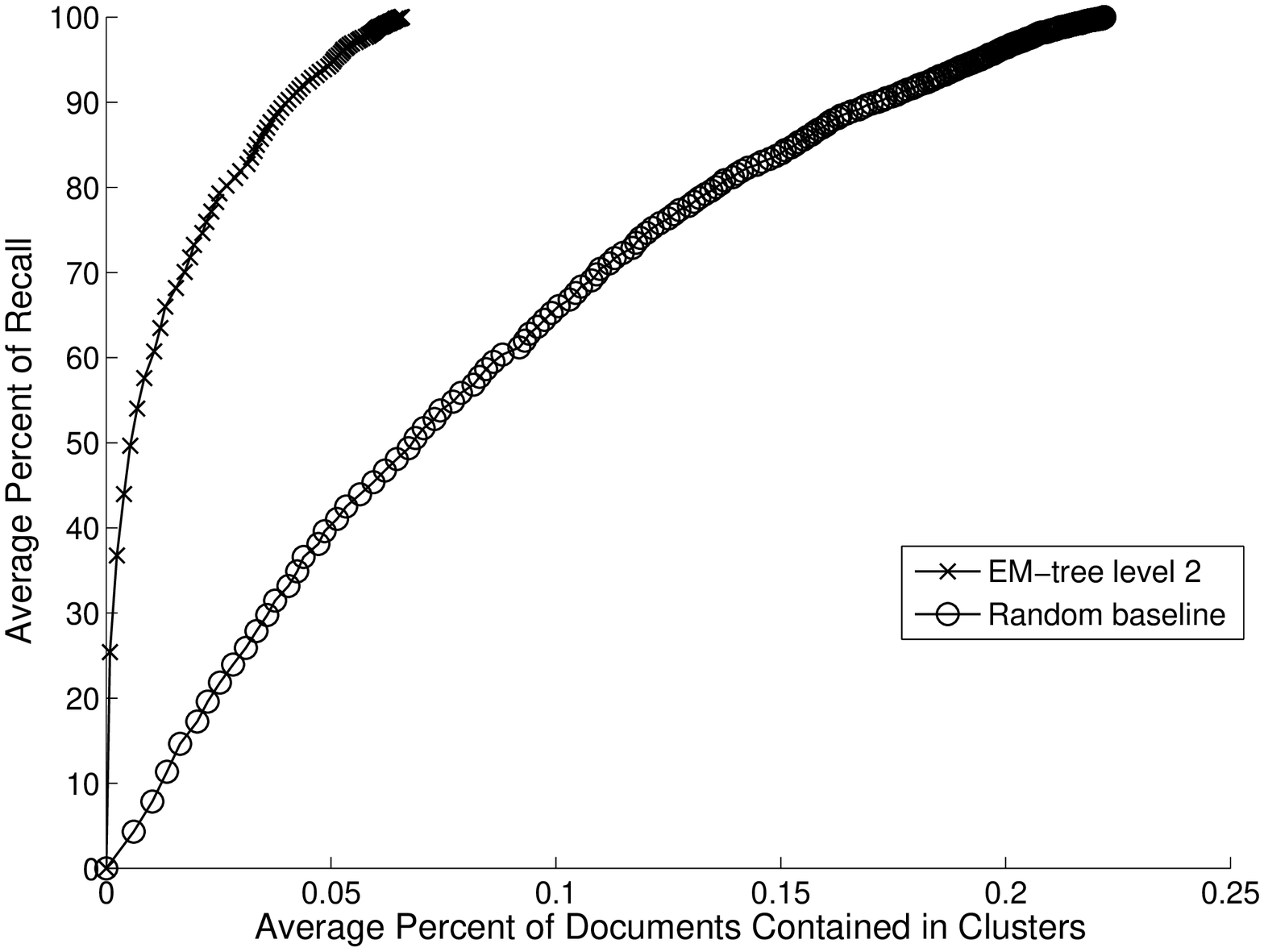}
\caption{ClueWeb09 -- 691,708 EM-tree level 2 clusters}
\label{fig:emtree_691708}
\end{minipage}
\end{figure*}

We generate cumulative recall plots for the three methods using the oracle collection selection approach. Each of the 3 approaches appear in separate plots in Figures \ref{fig:kulkarni_callan_1000}, \ref{fig:emtree_1000}, and \ref{fig:emtree_691708}. We place most relevant documents first using the oracle ranking process. We visit clusters in descending order of recall. The average percentage of total documents in the collection contained in the first $n$ clusters is display along the x-axis. The y-axis is the average percentage of recall included in the first $n$ clusters when averaged over all topics. Visiting a cluster produces a mark on the curve. So, after seeing 2 marks, 2 clusters have been visited in the order specified by the oracle ranking.

Each clustering has a unique random baseline created where the cluster size distribution matches that of each clustering solution. In all cases, there is a difference between the random baseline and the clustering solution. Useful learning has occurred. The baseline normalization removes any effect of random chance or ineffective cluster size distributions. Therefore, all three clustering solutions support the cluster hypothesis that relevant documents tend to cluster together.

\begin{figure*}
\begin{minipage}[b]{0.33\linewidth}
\includegraphics[width=\columnwidth]{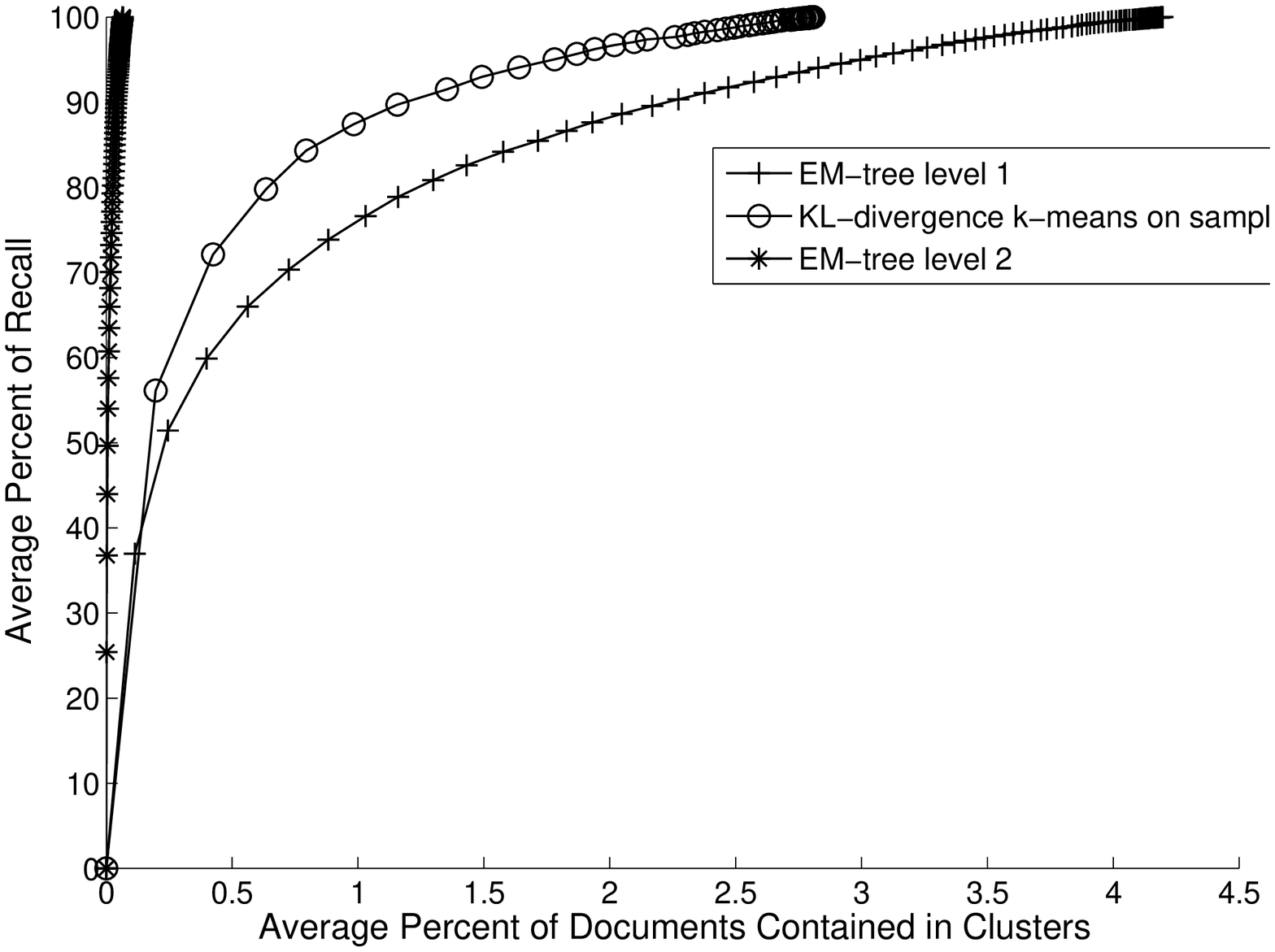}
\caption{ClueWeb09 -- Comparing all approaches}
\label{fig:all}
\end{minipage}
\begin{minipage}[b]{0.33\linewidth}
\includegraphics[width=\columnwidth]{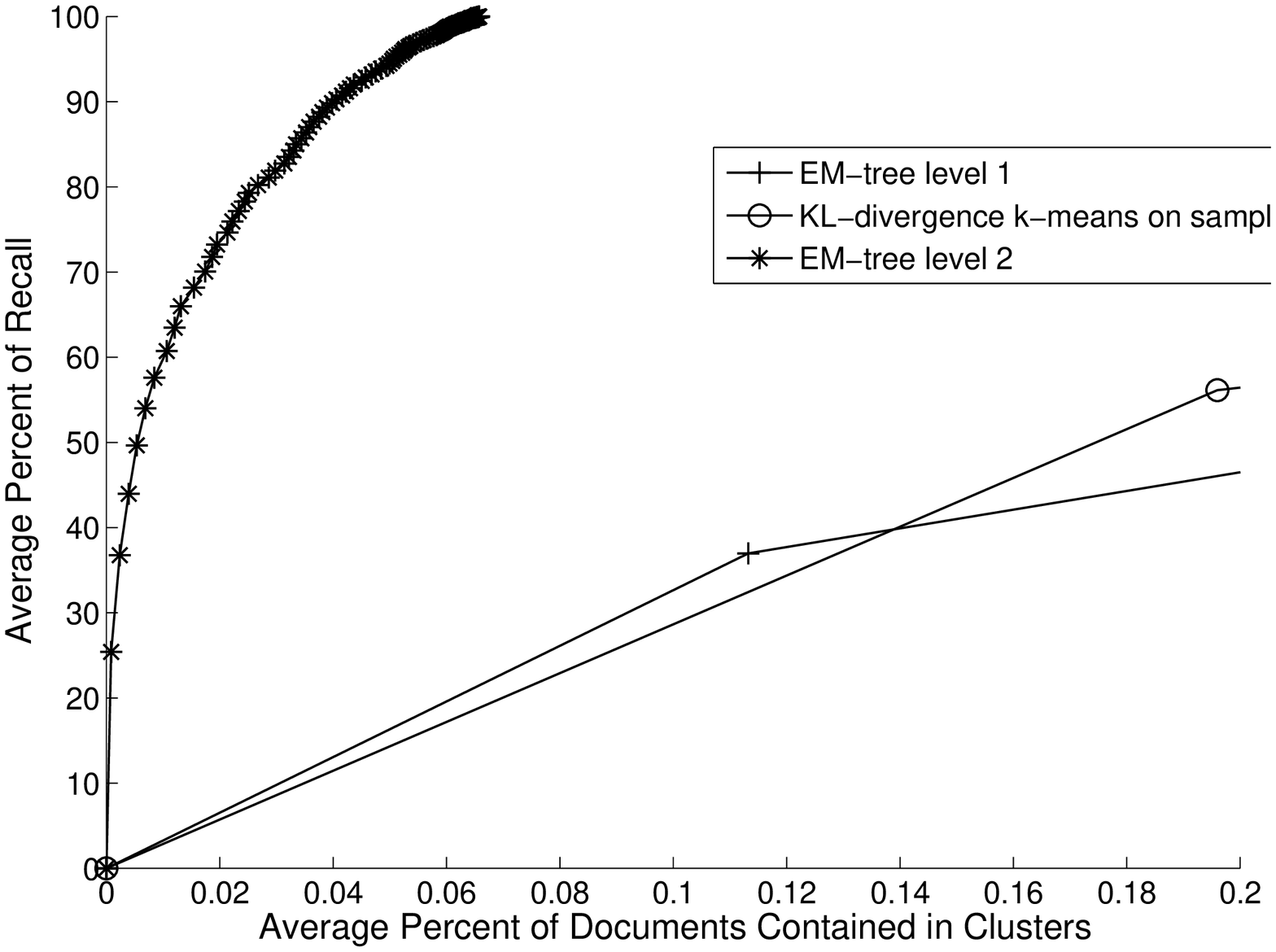}
\caption{ClueWeb09 -- Comparing all approaches -- Zoomed}
\label{fig:all_zoom}
\end{minipage}
\begin{minipage}[b]{0.33\linewidth}
\includegraphics[width=\columnwidth]{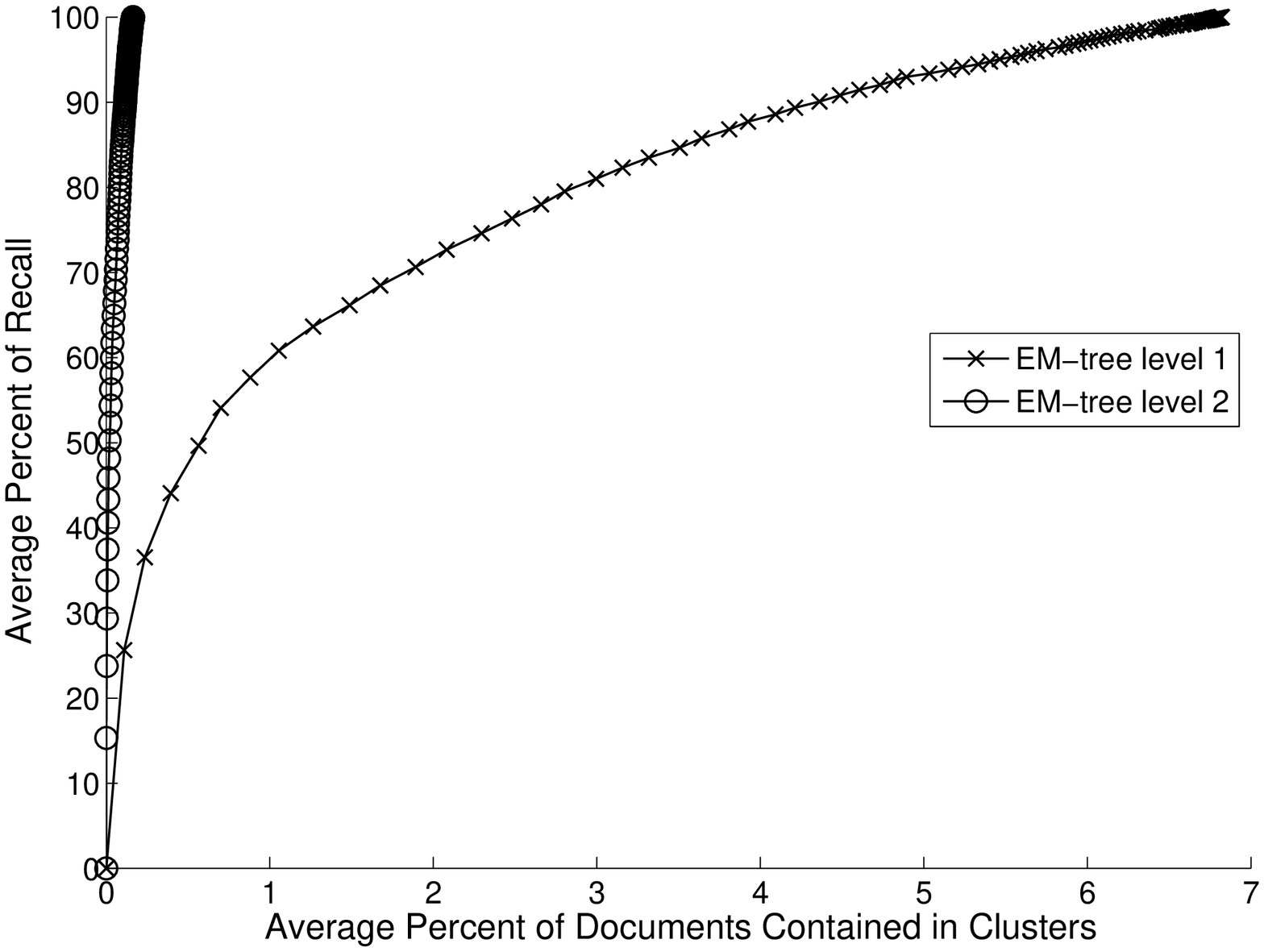}
\caption{ClueWeb12 -- Comparing all approaches}
\label{fig:all_clueweb12}
\end{minipage}
\end{figure*}

In comparison to the 1000 clusters produced by the first level of EM-tree in Figure \ref{fig:emtree_1000}, the KL-divergence k-means approach in Figure \ref{fig:kulkarni_callan_1000} clearly groups relevant documents better. It achieves total recall after 2.81\% of the collection is visited, whereas the EM-tree approach does not achieve this until 4.19\%. Figure \ref{fig:all} highlights difference between the two approaches. All three approaches are plotted on one graph without their baselines. We note that this is not a surprising result -- k-means is superior to EM-tree in terms of quantization distortion rates. However, this is a trade off as it is also much more computationally intensive. If a particular application requires 1000 clusters then, the KL-divergence with k-means approach is clearly better at grouping relevant documents together. A 500,000 document sample provides adequate statistics for learning 1000 clusters. However, suppose that fine granularity clustering is sought after and applying k-means recursive clustering to the entire collection is necessary. We clustered 500,000 documents into 1000 clusters using a sparse vector approach using k-means, and it took approximately ten hours using a very fast implementation. If each of these 1000 clusters are clustered into another 1000 clusters using the recursive k-means approach, then each subset requires another 500,000 document sample to learn another 1000 subclusters. As there are 1000 partitions, taking ten hours per partition, it would take another 10,000 hours to produce a similar number of clusters as EM-tree, making this k-means based approach infeasible.

EM-tree can produce many more clusters in the second level that group relevant documents much more tightly. In Figures \ref{fig:all} and \ref{fig:all_zoom}, the second level of EM-tree reaches total recall after 0.06\% of the collection has been visited. Because these clusters still group relevant documents together, and support the cluster hypothesis, we also conclude they are highly topical. Relevant documents for a given query are related to the same topic because they satisfy the same information need. Many applications can benefit from identifying small clusters of highly topical documents. This is discussed further in Section \ref{sec:applications}.

The random baseline for the second level of EM-tree looks surprisingly effective. It achieves total recall after only visiting 0.22\% of the collection. It is not unexpected because this baseline is close to the worst case possible, where each relevant document for a given query appears in a separate cluster. There are on average 80 relevant documents per query for the 148 queries. So the oracle returns an average of 80 clusters per query to achieve 100\% recall. Indeed, the first cluster in the random baseline contains 4.3\% or 3.44 relevant documents on average. It decreases to below the average of one document per cluster by the 15th cluster. Some clusters still contain more than one relevant document on average because the cluster size distribution in the random baseline matches that of clustering. Large clusters can receive more relevant documents due to their size.

In contrast, the first EM-tree cluster in the second level contains 25\% or 20 relevant documents on average. Additionally, the first EM-tree cluster contains 450 documents on average, whereas the first random baseline cluster contains 30,000 documents on average. We calculate precision by dividing the number of relevant documents by the number of documents in the cluster. The first EM-tree cluster has a precision of 4.44\% on average, whereas the first random baseline cluster has a precision of 0.011\% on average. It is some 400 times worse.

Furthermore, clusters in the random baseline contain a mixture of relevant documents with mostly random off topic documents. It makes it much harder for a non-oracle ranker to find them using vocabulary statistics. There are on average 3.4 documents mixed with 30,000 random other off-topic documents in the first cluster in the random baseline.

The plots of EM-tree level 1 and 2 in Figures \ref{fig:emtree_1000} and \ref{fig:emtree_691708} look very similar when the x-axis scale is ignored. It indicates that the same relationships between the random baseline and the clustering exist in both levels of the EM-tree. While the same relationships exist, the 2nd level contains 691 times more clusters. The cluster hypothesis holds in the same way, even though the size of the clusters is reduced by almost three orders of magnitude. It verifies that there is no significant loss of fidelity in fine granularity clustering.

We conducted the same experiments on ClueWeb12. We were unable to compare the KL-divergence approach \cite{Kulkarni2010} as clusters have not been made available for this recent collection. However, the same trends for the EM-tree emerge as shown in Figure \ref{fig:all_clueweb12}. All the corresponding plots were highly similar but were left out for brevity. The relevance judgments from the 2013 TREC Web Track have much deeper pooling, increasing our confidence in the results.

\subsection{Spam Based Evaluation}

We have performed another evaluation using external knowledge in the form of spam scores. These spam scores have been produced by an approach by Cormack et. al. \cite{Cormack2011} using a supervised learning approach that combines training from 4 different labeled spam datasets. It quantizes the spam scores into 100 values. A score of 99 is the least spammy and 0 the most. For instance, a document score of 99 indicates that 99\% of the collection is more spammy than the said document. These spam scores have shown to be useful for improving search quality in ad hoc retrieval \cite{Cormack2011}.

If a perfect document clustering exists with respect to spam, it places documents with the same spam score in the same cluster. We have evaluated clusters by taking the average spam score for documents in the cluster. Clusters are sorted and traversed by descending spam score. We then observe the percentage of the documents contained in the clusters visited so far. Again, we create a random baseline for each clustering solution. As expected, almost all clusters average out to a spam score of 50. We display the best possible oracle solution by ordering documents according to their spam score; i.e. all documents with a score of 99, then 98 and so on. It is a straight line where the first ranked cluster contains all documents with a spam score of 99, then those with a spam score of 98, and so forth. Therefore, the closer an actual clustering solution comes to this straight line from the top left to the bottom right, the better the clustering is with respect to the spam scores.

\begin{figure*}
\begin{minipage}[b]{0.33\linewidth}
\includegraphics[width=\columnwidth]{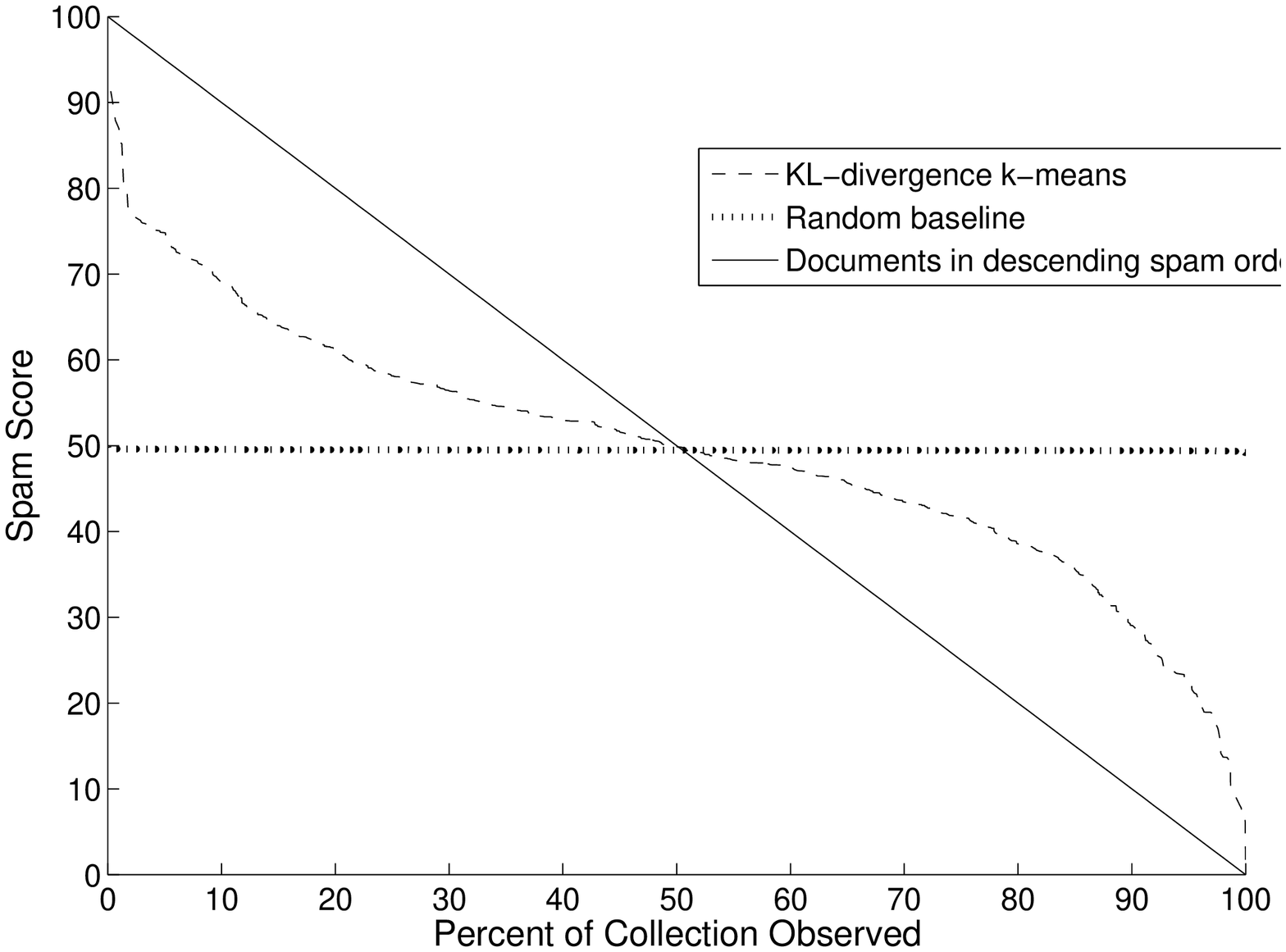}
\caption{ClueWeb09 -- 1000 KL-divergence k-means clusters}
\label{fig:spam_clueweb09_kulkarni_1000}
\end{minipage}
\begin{minipage}[b]{0.33\linewidth}
\includegraphics[width=\columnwidth]{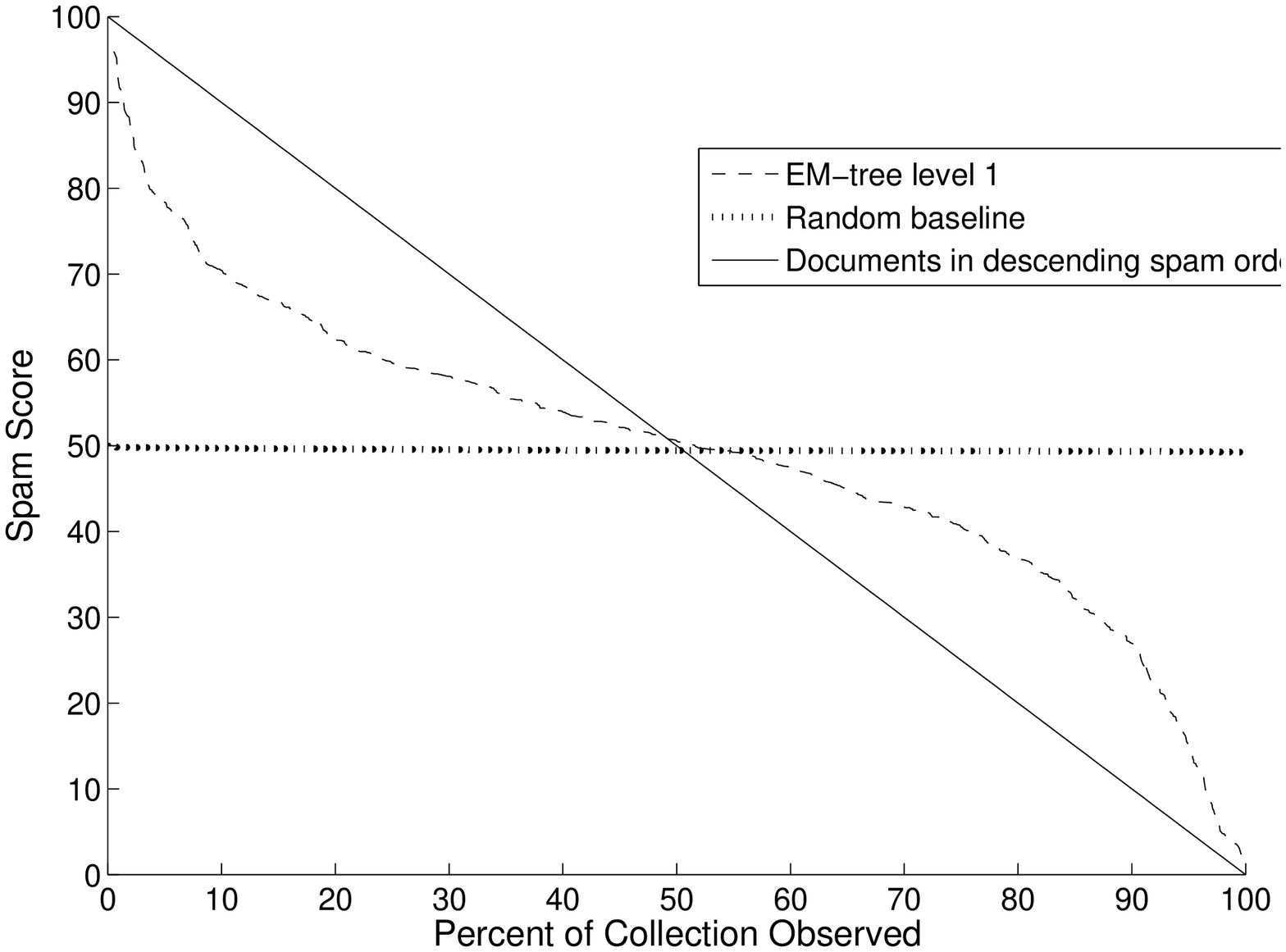}
\caption{ClueWeb09 -- 1000 EM-tree level 1 clusters}
\label{fig:spam_clueweb09_emtree_1000}
\end{minipage}
\begin{minipage}[b]{0.33\linewidth}
\includegraphics[width=\columnwidth]{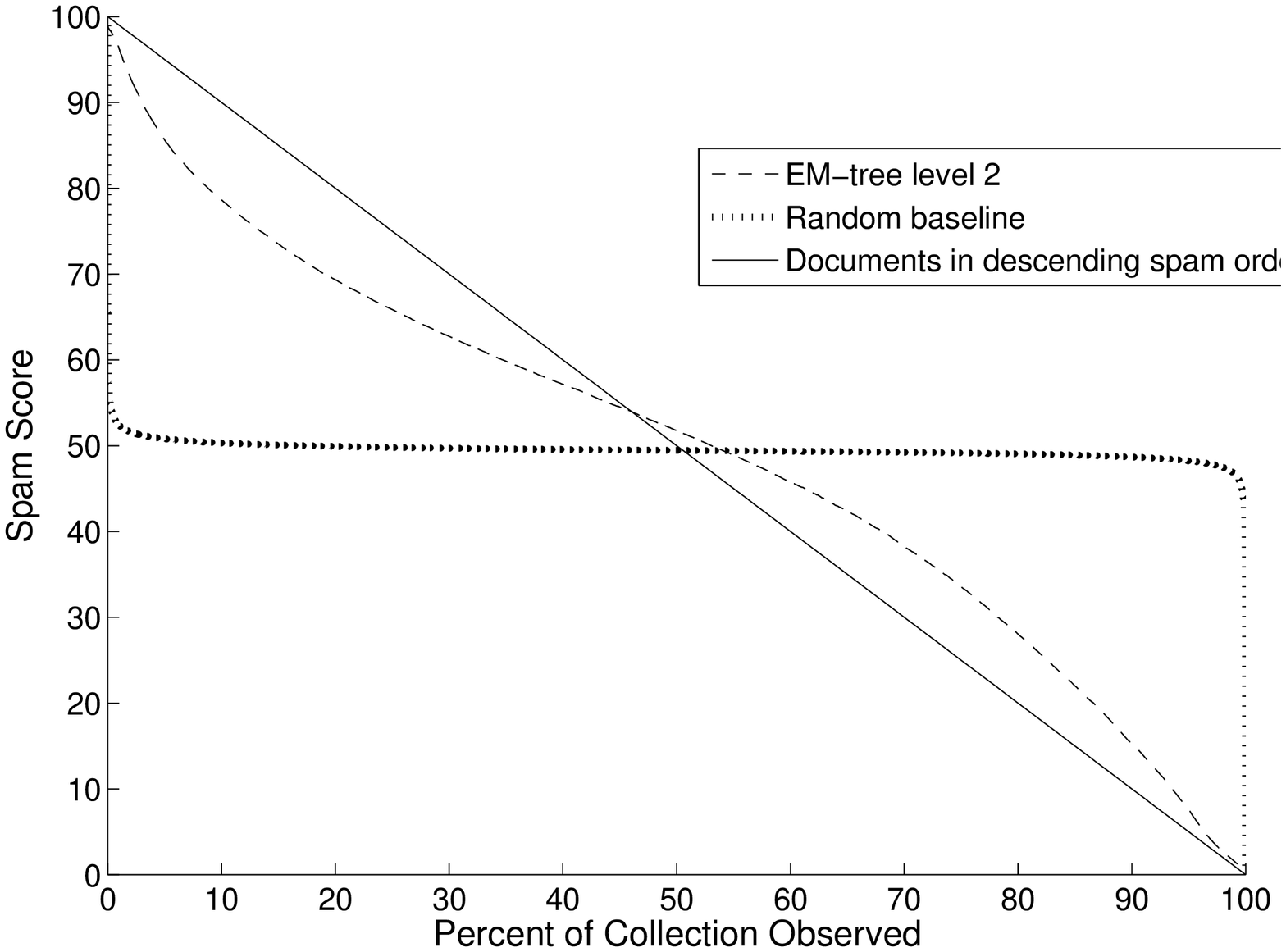}
\caption{ClueWeb09 -- 691708 EM-tree level 2 clusters}
\label{fig:spam_clueweb09_emtree_691708}
\end{minipage}
\end{figure*}

\begin{figure*}
\begin{minipage}[b]{0.33\linewidth}
\includegraphics[width=\columnwidth]{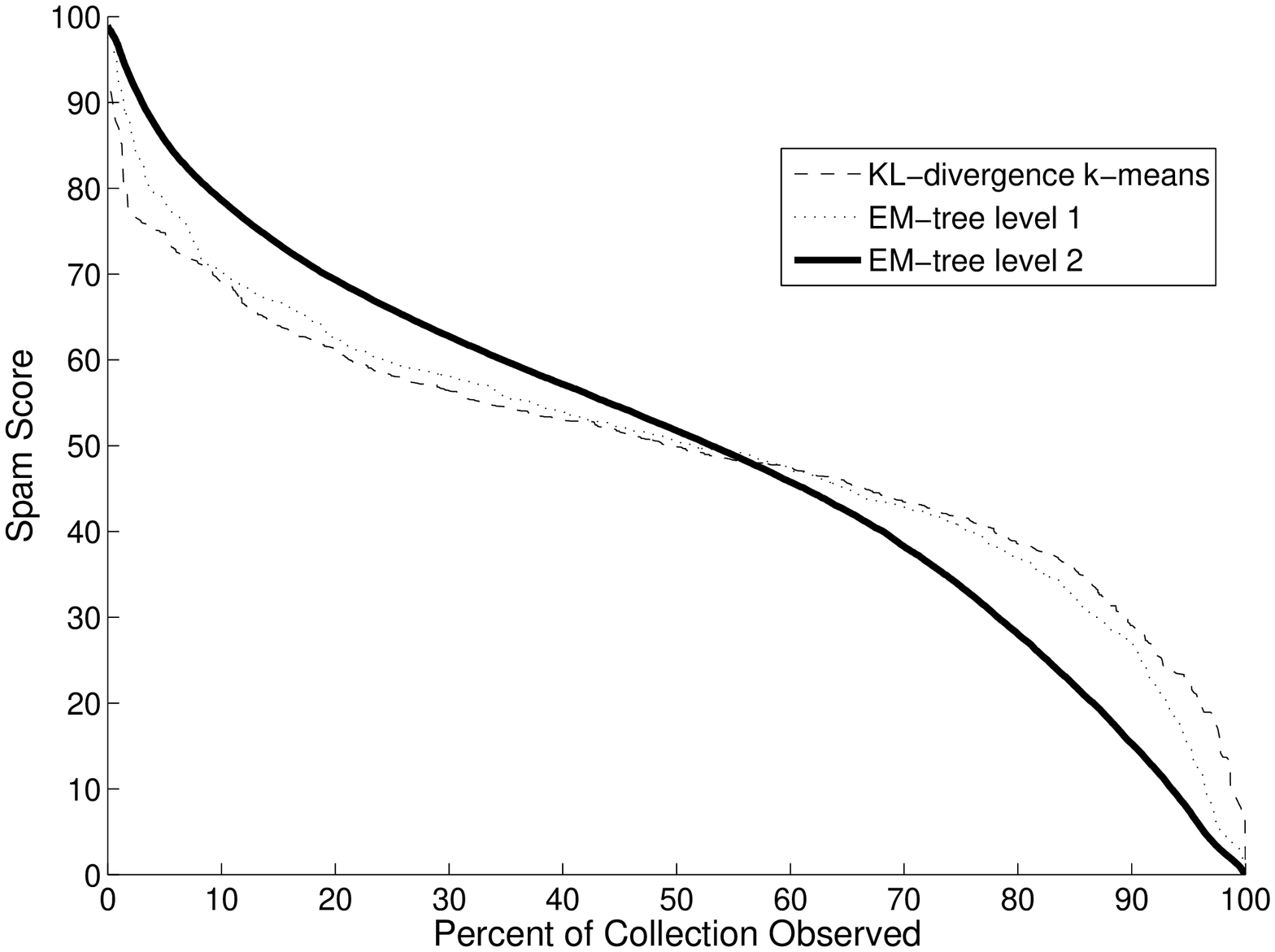}
\caption{ClueWeb09 -- Comparing all approaches}
\label{fig:spam_clueweb09_emtree_both}
\end{minipage}
\begin{minipage}[b]{0.33\linewidth}
\includegraphics[width=\columnwidth]{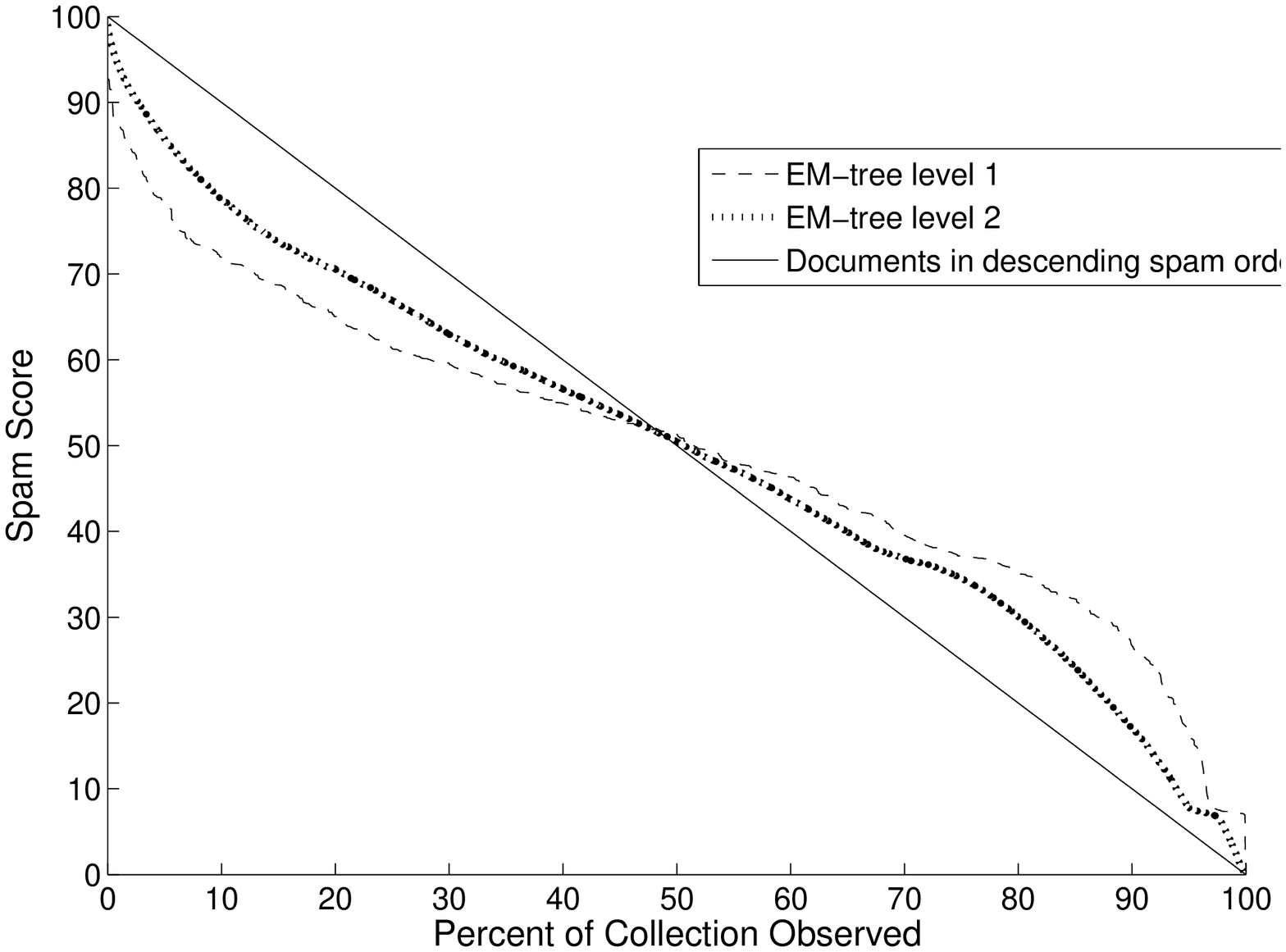}
\caption{ClueWeb12 -- Comparing all approaches}
\label{fig:spam_clueweb12_emtree_both}
\end{minipage}
\end{figure*}

We evaluated the same clustering solutions as in Section \ref{sec:relevance}. These are KL-divergence k-means \cite{Kulkarni2010} and the two clusterings produced by level 1 and 2 of the EM-tree. For the ClueWeb09 collection, the results of this experiment can be seen in Figure \ref{fig:spam_clueweb09_emtree_both}. Each of the clusterings have been plotted separately with random baselines and optimal document ordering in Figures \ref{fig:spam_clueweb09_kulkarni_1000}, \ref{fig:spam_clueweb09_emtree_1000} and \ref{fig:spam_clueweb09_emtree_691708}.

The signature-based approach with EM-tree better groups documents according to spam score than the KL-divergence k-means approach when producing 1000 clusters. It is the opposite result to using ad hoc relevance judgments where KL-divergence k-means was superior. The finer grained clustering in the 2nd level of EM-tree allows more pure clusters with respect to spam.

For ClueWeb12, we have only created the final plot comparing all approaches for brevity in Figure \ref{fig:spam_clueweb12_emtree_both}. We omitted the random baselines because they behave in exactly the same manner. The plots look very similar for ClueWeb12 and ClueWeb09.

\section{Applications}
\label{sec:applications}

Recent results in computer vision demonstrate the usefulness of scalable algorithms for unsupervised feature learning \cite{le_building_2011, coates_emergence_2012}. The EM-tree algorithm using signatures could be advantageous in computer vision applications where the generation of many clusters from large amounts of data is desirable.

Near duplicate object detection is another application for large scale fine-grained clustering, particularly for eliminating near-duplicate web pages. In fact, we found many near duplicate pages in ClueWeb 2012.

Most existing cluster based re-ranking and query expansion approaches use fine-grained clusters of the top k results returned by a search engine. It limits the analysis to documents returned by the search engine, and usually it is based solely on the keywords in the query. With fine-grained clusters of web-scale collections made available by the EM-tree algorithm, re-ranking and query expansion of documents that do not contain the keywords becomes a possibility. Common cluster membership associates documents based on other keywords that appear in relevant documents. Such documents do not necessarily contain the query keywords. Another area of information retrieval, collection selection, is apt for the application of our approach. Kulkarni and Callan \cite{Kulkarni2010} have demonstrated the effectiveness of selective search using collection selection and 1000 clusters of ClueWeb09. It only visits the first few clusters per query. Our results indicate that relevant results cluster in the same way, even when there are almost three orders of magnitude more clusters.

The clustering solutions produced by EM-tree may be useful for scaling classification applications. Nearby points contained in a few fine-grained clusters can be used to build a classifier. Alternatively, clusterings can be used for classification directly.

Sub-document clustering is now a possibility for large-scale document collections where splitting documents into fragments creates even more objects to cluster. It makes sense to split larger documents since such documents probably contain multiple sub-topics. However, splitting documents in web-scale collections certainly push it well beyond the capability of standard commodity hardware. However, with EM-tree it may be possible to consider such clustering of document fragments.

\section{Conclusion}
\label{sec:conclusion}

In this paper, we presented solutions for two major problems in web-scale document clustering -- scalable and efficient document clustering and evaluation of cluster validity where categorical labeling of collections is unavailable and unfeasible.

The proposed EM-tree algorithm can cluster hundreds of millions of documents into hundreds of thousands of clusters on a single 16 CPU core machine in under 24 hours. It is standard hardware available to organizations of all sizes. To the best of our knowledge, clustering on a single machine at this scale has not been reported in the literature. To the contrary, the few published attempts at this scale have used high-performance computing resources, well beyond reach of most organizations. These attempts were also on lower dimensional non-document data. For document clustering, this is far beyond any examples we have been able to find. The largest scale approach we could find sampled 500,000 documents to produce 1,000 clusters. It then maps the 500 million documents onto these clusters \cite{Kulkarni2010}. The closest comparison we could find is a method that clusters the same order of magnitude of images into a similar order of magnitude of clusters using 2000 CPUs \cite{Liu2007,wang2013duplicate}. However, this approach does not cluster documents that are sparse and high dimensional. Another approach parallelized k-means using 16,000 CPUs on a super computer to produce 1,000 clusters of 1 billion much lower dimensional weather data \cite{Bisgin2008}.

We presented two novel evaluations using ad hoc relevance and spam classifications to assess the validity of clusters where no category labels are available. This evaluation demonstrated that the fine-grained clustering created by EM-tree led to higher quality clusters. Additionally, we expect this approach to clustering to be applicable to many different types of web scale data emerging from sources such as images, sound recordings, sensors, positioning systems, genome research, marketing data and many more fields.

{\small
\bibliographystyle{plain}

}
\end{document}